\documentclass{epl}
\usepackage{amssymb}

\institute{
  Yerevan State University, 1 Alex Manoogian Str., Yerevan, 375025,
 Armenia}
\def\bea{\begin{eqnarray}}
\def\eea{\end{eqnarray}}
\def\be{\begin{equation}}
\def\ee{\end{equation}}

\input{tcilatex}
\begin{document}

\title{A phenomenological theory of phase transitions in high~$T_c$~
superconductors }
\author{D. M. Sedrakian}
\maketitle

\begin{abstract}
A phenomenological theory is suggested to explain the experimentally
discovered ``paramagnetic peculiarity'' and the unconventional change of the
energetic state of a layers of high~$T_c$~ of Y-Ba-Cu-O in the vicinity of
the phase transition. The physical conditions are found under which these
peculiarities are revealed. It is shown that the suggested theory
qualitatively describes the experimental data.
\end{abstract}


Despite the substantial success in the investigations of the high~$T_c$~
superconductors, the microscopic nature of the superconducting state in
these materials has not been understood to date \cite{1}. In recent years
promising attempts of experimental investigations of the superconducting
phase transition were made in high~$T_c$~ Y-Ba-Cu-O \cite{2,3}. In this work
a new ``paramagnetic'' peculiarity and unconventional change in the
energetic state of layer of high~$T_c$~ superconductor in the vicinity of
the phase transition is discovered. This success was achieved due to the
construction of a highly sensitive magnetometer, which permits to measure
very small changes $\Delta\lambda$ of the penetration depth of the magnetic
field with frequency of the order several MHz in a sample of a flat high~$%
T_c $~ superconductor. The experimental set-up permits to measure changes $%
\Delta\lambda$ of absolute magnitude of the order $\Delta\lambda\sim 1-3$ A
with relative accuracy $\Delta\lambda/\lambda \sim 10^{-6}$ \cite{2}. The
first of these peculiarities shows an increase of the penetration depth of
the order of few micrometers when the temperature is decreased in the
vicinity of the phase transition, prior to its decrease from the value of $%
\delta$ of the order of hundreds of micrometers up to the London penetration
depth $\lambda$ of the order of a few micrometer \cite{2}. The second
peculiarity is manifested in an increase of the energy of the sample with
decreasing temperature prior to the well-known decrease due to the
transition to the superconducting state \cite{3}.

The aim of this work is to show within the framework of a phenomenological
theory of superconductivity that peculiar dependence of the the penetration
depth on the temperature can be explained within an approach, which suggests
a specific behavior of the Cooper pairs. We will show as well, that the
account of the Coulomb interaction in a solid-state plasma, which is
composed of normal ions a and normal and superconducting electrons, can
explain the observed absorption of the energy of the electromagnetic field
during the superconducting phase transition \cite{3}.

Section 2 derives an expression for the penetration depth $\lambda$ and the
conditions are obtain under which the maximum of the $\lambda$ can be at $%
T=85.4$ K, as observed in ref. \cite{2}. In Section 3the electrostatic
potential is found for a charged particle $Ze$ with account of the screening
of the surrounding plasma. In the closing Section 4 the change in the free
energy of the sample with decreasing temperature is computed. Here we show
that the energy of the sample as a function of temperature shows a maximum
and only beyond the maximum one observes the familiar temperature dependence
of the difference in the free-energies of the superconducting and normal
phases.

\section{The penetration depth of the electromagnetic field in a high~$T_c$~
superconductor}

Suppose that in the superconductor the electric current is a sum
of the normal and superconducting currents, $\vec j_n$ and $\vec
j_s$. These currents are related to an external electric field
$E\sim exp(-i\omega t)$ by the following relations
\begin{eqnarray}  \label{2}
\vec j_n &=& \frac{e^2\tau}{m} \frac{n_n}{1-i\omega\tau}\vec E, \\
\vec j_s &=& i\frac{e^2n_s}{m\omega} \alpha\vec E,
\end{eqnarray}
where $n_n$ and $n_s$ are the densities of the normal and superconducting
electrons, respectively, $\tau$ is the mean-free-flight time of a normal
electron, $e$ and $m$ are the charge and the mass of electron and $\alpha$
is the fraction of the Cooper pairs which are involved in the supercurrent.

Our main assumption is that when $T<T_c$ not all the Cooper pairs
participate in the current, i.e. $\alpha <1$, and with a decrease of the
temperature, $\alpha $ increases and, already for the temperatures which
correspond to the fast decrease of the penetration depth, tends to unity.
Such a behavior of $\alpha $ can be understood in the frame of nonlocal
theory of Pippard [4], according which in the well-known London's connection
Eq.(2) the coefficient

\smallskip
\[
\alpha =\frac \xi {\xi _0}=\frac 1{\xi _0/l+1}
\]
have to appear. Here $\xi _0$ is the coherent lenght of Cooper pairs and $l$
is the lenght of free path of electrons. Near the phase transition $%
T\lesssim T_c$ $\xi _0\gg l,$ hence $\alpha \approx l/\xi _0\ll 1$. With
decrease of temperature the superconducting correlations between electrons
bring to the decrease of $\xi _0$ and increase of $l$, hence $\xi _0/l$ will
become less than unity and $\alpha $ will tend to one. When $\alpha =1$ the
dependence of $\lambda $ on $T$ corresponds to the familiar dependence $%
\lambda =\lambda (T)$, which follows from the theory of the
superconductivity.The experimental measurement of the temperature dependence
of the penetration depth $\lambda (T)$ would allow to find form of the
function $\alpha (T)$.

Upon substituting the net current $\vec{j}=\vec{j}_s+\vec{j}_n$ in the
Maxwell equation
\begin{equation}
\mathrm{curl}\vec{B}=\frac{4\pi }c\vec{j}  \label{3}
\end{equation}
and acting with the curl operator on both sides of Eq. (\ref{3}) and using
another Maxwell equation
\begin{equation}
\mathrm{curl}\vec{E}=-\frac 1c\frac{\partial \vec{B}}{\partial t},  \label{4}
\end{equation}
we finally find
\begin{equation}
\mathrm{curl~curl}\vec{B}=\frac{4\pi e^2n}{m_c^2}\left( \frac{i\omega \tau }{%
1-i\omega \tau }\frac{n_n}n-\alpha \frac{n_s}n\right) \vec{B}.  \label{5}
\end{equation}
If we assume that the magnetic field depends only on the distance $z$ from
the surface of the superconductor, then Eq. (5) within the bulk of a
superconductor $z>0$ assumes the form
\begin{equation}
\frac{d^2\vec{B}(z)}{dz^2}+k^2\vec{B}(z)=0,  \label{6}
\end{equation}
where
\begin{equation}
k^2=\frac{4\pi e^2n}{m_c^2}\left( \frac{i\omega \tau }{1-i\omega \tau }\frac{%
n_n}n-\alpha \frac{n_s}n\right) .  \label{7}
\end{equation}
The solutions of Eq. (\ref{6}) are of the from
\begin{equation}
\vec{B}(z)=\vec{B}_0e^{ikz-i\omega t},  \label{8}
\end{equation}
i.e. the penetration depth of an electromagnetic field is
\begin{equation}
\lambda (T)=\frac 1{\mathrm{Im}k}.  \label{9}
\end{equation}
Introducing the short hand notations
\begin{equation}
\lambda _0^2=\frac{mc^2}{4\pi e^2n},\quad \delta _0^2=\frac{mc^2}{2\pi e^2n}%
\frac{1+\omega ^2\tau ^2}{\omega \tau },\quad Z_0=\frac{\delta _0^2}{%
2\lambda _0^2}=\frac{1+\omega ^2\tau ^2}{\omega \tau },\label{10}
\end{equation}
we find
\begin{equation}
\lambda =\delta _0\Biggr[ \sqrt{\left( \frac{n_n}n\right) ^2+Z_0\left(
\alpha \frac{n_s}n+\frac{\omega ^2\tau ^2}{1+\omega ^2\tau ^2}\frac{n_n}%
n\right) ^2}+Z_0\left( \alpha \frac{n_s}n+\frac{\omega ^2\tau ^2}{1+\omega
^2\tau ^2}\frac{n_n}n\right) \Biggl]^{-1/2}.  \label{11}
\end{equation}
A compact form for the penetration depth (\ref{11}) is obtain by introducing
further abbreviation
\begin{equation}
\frac{n_n}n=x,\quad \frac{n_s}n=1-x,\quad \alpha Z_0=Z,\quad \beta =1-\frac{%
\omega \tau }Z;  \label{12}
\end{equation}
we find
\begin{equation}
\lambda (T)=\delta _0\left[ \sqrt{x^2+Z^2(1-\beta x)^2}+Z(1-\beta x)\right]
^{-1/2}.  \label{13}
\end{equation}
Since $\omega \tau \ll 1$ and the minimal value of $Z$ is of the order of
unity, we replace with high accuracy $\beta $ by 1. Then Eq. (\ref{13})
simplifies to the from
\begin{equation}
\lambda (T)=\delta _0\left[ \sqrt{x^2+Z^2(1-x)^2}+Z(1-x)\right] ^{-1/2}.
\label{14}
\end{equation}
As can be seem from Eq. (\ref{14}), for $T>T_c$, when $x=1$, the penetration
depth $\lambda $ is equal the depth of the skin effect $\delta _0$. With the
decreasing temperature, when $Z\simeq 1$, that $\lambda (T)$ increases and
passes through a maximum at $T=T_0$ and becomes equal $\delta _0$ at
temperature $T=T_1$. As follows from ref. \cite{2}, $T_0=85.4$ K, $T_1=85$
K, whereas $T_c=88.7$ K. The relative change of the penetration depth at the
maximum is $\Delta \lambda /\delta _0\simeq 5~10^{-3}$. Assuming that in the
temperature range $T_1\le T\le T_c$ $Z$ is constant, we find the maximal
value of $\lambda (T)$ from Eq. (\ref{14}). A simple calculation shows that
if the maximum of $\lambda (T)$ is located at the point $x_0$, then this
point is determined by the formula
\begin{equation}
x_0(2-x_0)=\left[ \frac 1{1+\Delta \lambda /\delta _0}\right]   \label{15}
\end{equation}
At the point were $\lambda (x)$ is again equal $\delta _0$ the following
relations hold
\begin{equation}
x_1=2Z^2-1,  \label{16}
\end{equation}
where
\begin{equation}
Z^2=\frac{x_0}{2-x_0}.  \label{17}
\end{equation}
Substituting in Eq. (\ref{15}) the value of $\Delta \lambda /\delta _0$, we
obtain from (\ref{15}) and (\ref{17}) $x_0=0.8586$ and $Z=0.8673$. If,
following the Gorter-Casimir theory, one assumes $x_0=(T_0/T_c)^4$, then
from the requirement that the maximum is at $T_0=85.4$ K, we obtain the
value of $T_c=88.7$ K, which coincides with the value of $T_c$ quoted in
ref. \cite{2}. If we require that $T_1=85$ K, then one needs to take the
value $Z=0.92$; this shows that there is an initial increase of $Z$ which
accompanies the drastic decrease $\lambda (T)$. The further decrease of $%
\lambda (T)$ with decreasing temperature, seen in the experiment \cite{2},
can be explained by the increase in $Z$, i.e. an increase of $\alpha $ up to
unity. A more detailed comparison between Eq. (\ref{14}) and the results of
ref. \cite{2} will be given in future.Of great interest is the derivation of
the function $\alpha =\alpha (T)$ from the microscopic theory.

\section{Electrostatic potential in the solid-state plasma at $T\to 0$}

As is well known from the plasma theory, the electrostatic potential of a
charged particle in plasma is screened by the distribution of charged
particles of opposite sign \cite{5}. The Fourier-component of the
longitudinal component of the dielectric constant $\varepsilon (0,k)$ when $%
\omega \to 0$ for mixture of ions, normal and superconducting electrons has
the following form
\begin{equation}
\varepsilon _l(k,0)=1+\frac 1{k^2a_i^2}+\frac 1{k^2a_e^2}+\frac 1{k^2a_s^2},
\label{18}
\end{equation}
where $a_i$, $a_e$ and $a_s$ are the Debye radii of ions, normal and
superconducting electrons, respectively. The radius $a_i$ can be found from
the formulae \cite{6}
\begin{equation}
a_i=\frac{\nu _i}{\Omega _i},\quad \nu _i=\sqrt{\frac EM}  \label{19}
\end{equation}
where $E$ is the mean kinetic energy, $M$ is the mass of the ion, $\Omega _i$
is the ion plasma frequency,
\begin{equation}
\Omega _i=\left( \frac{4\pi n_iZ_i^2e^2}M\right) ^{1/2}.  \label{20}
\end{equation}
Here $eZ_i$ and $n_i$ are the charge and the density of ions, respectively.
From Eqs. (\ref{19}) and (\ref{20}) we obtain for $a_i$ the following
expression
\begin{equation}
a_i=\sqrt{\frac{E_i}{4\pi e^2n}}  \label{21}
\end{equation}
where
\begin{equation}
E_i\ =E\frac n{n_iZ_i^2}.  \label{22}
\end{equation}
Considering the sample of the high~$T_c$~ superconductor as a solid, at low
temperatures for the mean kinetic energy of ion vibrations one can use the
following expression
\begin{equation}
E=\frac{k_BT}2D(\frac \theta T),  \label{23}
\end{equation}
where $D(\frac \theta T)$ is the Debye function, $\theta $ is the Debye
temperature. In analogous manner one can find the Debye radius of the normal
electrons
\begin{equation}
a_e=\sqrt{\frac{\varepsilon _F}{4\pi e^2n_n}},  \label{24}
\end{equation}
where $\varepsilon _F$ is the electron Fermi-energy. The role of the Debye
radius of the superconducting electrons, as shown in \cite{7}, plays the
London penetration depth $\lambda $, i.e.
\begin{equation}
a_s\equiv \lambda =\sqrt{\frac{mc^2}{4\pi e^2n_s\alpha }},  \label{25}
\end{equation}
where $n_s$ is the density of the superconducting electrons. If we define
\begin{equation}
a^{-1}=\sqrt{\frac 1{a_i^2}+\frac 1{a_e^2}+\frac 1{a_s^2}},  \label{26}
\end{equation}
then the Fourier-component of the electric statical potential $\phi _k$ will
take the form \cite{6}
\begin{equation}
\phi _k=\frac{eZ_i/\epsilon _0}{k^2\varepsilon _l(0,k)}=\frac{eZ_i/\epsilon
_0}{k^2+a^{-2}},  \label{27}
\end{equation}
where $\epsilon _0$ is the dielectric constant of the sample under
consideration. Finally, for the electrostatical potential of a charge
particle $eZ_i$, we can write the following expression
\begin{equation}
\phi _i(r)=\int \phi _ke^{i\vec{k}\cdot \vec{r}}\frac{d\vec{k}}{(2\pi )^3}=%
\frac{eZ_i}{\epsilon _0r}e^{-r/a},  \label{28}
\end{equation}
where $r$ is the distance to the charge $eZ_i$, while $a$ is determined from
the Eqs. (\ref{21})-(\ref{26}).

\section{The Coulomb interaction energy of solid-state plasma}

The energy of the Coulomb interaction between the charged particles in
plasma is of the form \cite{5}
\begin{equation}
W=\frac 12V\sum_iZ_ien_i\phi _i^{\prime }.  \label{29}
\end{equation}
Here $n_i$ is the density of the charged particles, $\phi _i^{\prime }$ is
the potential created by all charged particles at the point $r=0$, except
the field of the particle with the charge $eZ_i$, located at that point and $%
V$ is the volume. Expand the potential $\phi _i(r)$, Eq. (\ref{28}), in the
vicinity of the point $r=0$:
\begin{equation}
\phi _i(r)=\frac{eZ_i}{\epsilon _0r}\left( 1-\frac ra\right) =\frac{eZ_i}{%
\epsilon _0r}-\frac{eZ_i}{\epsilon _0a}.  \label{30}
\end{equation}
Consequently, $\phi _i^{\prime }$ can be identified with the second term on
the left-hand-side of the expression (\ref{30}). Substituting this
expression in Eq. (\ref{29}), we obtain the energy per unit volume
\begin{equation}
\varepsilon _c=\frac WV=-\frac{e^2}{2\epsilon _0}\sum_iZ_i^2n_i/a.
\label{31}
\end{equation}
Using the Eqs. (\ref{21})-(\ref{26}) we can compute the quantity $a$. Taking
into account the smallness of the quantities $E^{\prime }/\epsilon _F$ and $%
E^{\prime }/mc^2$ compared to unity, we rewrite $\varepsilon _c$ as
\begin{equation}
\varepsilon _c=-\frac{e^2}{2\epsilon _0}\sum_i\left( \frac{4\pi e^2n}{E_i}%
\right) ^{1/2}Z_i^2n_i\left[ 1+\frac{E_i}{2\varepsilon _F}\frac{n_n}n+\frac{%
E_i\alpha }{2mc^2}\frac{n_s}n\right] .  \label{32}
\end{equation}
The energy density of the Coulomb interaction $\varepsilon _c$ in the case
where electrons are normal follows from (\ref{32}) when $n_n/n=1$ and $n_s=0$%
, i.e.
\begin{equation}
\varepsilon _c=-\frac{e^2}{2\epsilon _0}\sum_i\left( \frac{4\pi e^2n}{E_i}%
\right) ^{1/2}Z_i^2n_i\left[ 1+\frac{E_i}{2\varepsilon _F}\right] .
\label{33}
\end{equation}
The gain in the energy density of the sample due to the coulomb interaction
is
\begin{equation}
\Delta \varepsilon _c=\varepsilon _c-\varepsilon _c^n=\frac{e^2n}{4\epsilon
_0\varepsilon _F}\sum_i(4\pi e^2nE_i)^{1/2}\frac{Z_i^2n_i}n\left( 1-\frac{%
\alpha \varepsilon _F}{mc^2}\right) \frac{n_s}n.  \label{34}
\end{equation}
If we take into account that $\alpha <1$ and $\varepsilon _F/mc^2\ll 1$ and
substitute for $E_i$ the expression (\ref{22}) in (\ref{34}) then we obtain
\begin{equation}
\Delta \varepsilon _c=\frac 1{\epsilon _0}\sum_i\left( \frac{\pi Z_i^2n_i}{4n%
}\right) ^{1/2}\frac{(e^2n)^{3/2}}{\varepsilon _F^{1/2}}\left( \frac
E{\varepsilon _F}\right) ^{1/2}\frac{n_s(T)}n.  \label{35}
\end{equation}
If we substitute for $E$ the expression (\ref{23}) and use the asymptotical
expression for the function $D(T/\theta )$ in the limit $T/\theta \ll 1$, we
obtain
\begin{equation}
\Delta \varepsilon _c=A\left( \frac T{T_c}\right) ^2\frac{n_s(T)}n,
\label{36}
\end{equation}
where
\begin{equation}
A=\left( \frac{\pi ^5}{40}\right) ^{1/2}\frac{(e^2n)^{3/2}}{\epsilon _0}%
\frac{(k_B\theta )}{\varepsilon _F}^{1/2}\left( \frac{T_c}\theta \right)
^2\sum_i\left( \frac{Z_i^2n_i}n\right) ^{1/2},
\end{equation}
and $T_c$ is the temperature of the superconducting phase transition.
According to the phenomenological Gorter-Casimir theory
\begin{equation}
\frac{n_s(T)}n=1-\left( \frac T{T_c}\right) ^4,  \label{38}
\end{equation}
which can be used in Eq. (\ref{36}) to eliminate $n_s$. The correction to
the free energy density of the normal state $\Delta f_c$ can be obtained
from $\Delta \varepsilon _c$ by integrating the thermodynamical relation $%
\Delta \varepsilon /T^2=-(\partial /\partial T)(\Delta f/T)$
\begin{equation}
\Delta f_c=f_c-f_c^n=-T\int \frac{\Delta \varepsilon _c}{t^2}dt+C.
\label{39}
\end{equation}
The integration constant $C$ is determined from the requirement $\Delta
f_c(T_c)=0$. Substituting Eqs. (\ref{36}) and (37) into Eq. (\ref{39}) and
carrying out the integration, we finally find
\begin{equation}
\Delta f_c=0.2A\left\{ 4-\left( \frac T{T_c}\right) ^2\left[ 5-\left( \frac
T{T_c}\right) ^4\right] \right\} .  \label{40}
\end{equation}
To obtain the total change in the free energy density due to the
superconducing phase transition within the sample, we need to add to $\Delta
f_c$ the change of free energy density due to the correlation interaction of
the Cooper pairs
\begin{equation}
\Delta f_s=-\frac{H_c^2(0)}{8\pi }\left[ 1-\right( \frac T{T_c}\left)
^2\right] ^2,  \label{41}
\end{equation}
where $H_c(0)$ is the critical magnetic field at $T=0$. The net change in
the free energy density is
\begin{equation}
\Delta f=\Delta f_c+\Delta f_s=0.2A\left[ 4-y(5-y^2)\right] -b(1-y)^2,
\label{42}
\end{equation}
where $b=H_c(0)^2/8\pi $ and $y=(T/T_c)^2$.

One should note that the change in the free energy in superconductors is
commonly given by the second term in Eq. (\ref{42}), consequently, as the
temperature is decreased it decrases and reaches the value $-b$ at $T=0$. As
shown in the exprimental work \cite{3}, such a behaviour of the variation of
the free energy density is not realized in the experiment. The
experimnetally deduced dependence of $\Delta f$ on temperatrue qualitatively
conincides with the that predicted by (\ref{42}). With decreasing
temperature $\Delta f$ increases first reaching a maximum at $y_0$ and
dcreases subseqently going through zero at the point $y_1$. As the
experimnet shows the deviations of $y_0$ from the $y_1$ is small compared to
unity. To guarantee that this is the case it is sufficient to require $A\ll
b $. An estimate of the values of $A$ and $b$ shows that this condition can
be easily met. In the framework of the Ginzburg-Landau theory $\alpha n_s/n$
is the square of the modulus of the equilibrium value of the order parameter
$|\psi _e|^2$. If the change in the free-energy density in the
Ginzburg-Landau theory is written in the form
\begin{equation}
\Delta f(\psi ,T)=c(T)|\psi |^2+\frac{d(T)}2|\psi |^4,  \label{43}
\end{equation}
then in the framework of the present theory
\begin{equation}
|\psi _e|^2=-\frac{c(T)}{d(T)}=\alpha (1-x),\quad \Delta f=-\frac 12\frac{%
c^2(T)}{d(T)}.  \label{44}
\end{equation}
Thus, the functions $c(T)$ and $d(T)$ can be determined from the
measurements $\lambda (T)$ and $\Delta f(T)$ according the the following
relations
\begin{equation}
c(T)=\frac{2\Delta f}{\alpha (1-x)}\quad d(T)=-\frac{2\Delta f}{\alpha
^2(1-x)^2}  \label{45}
\end{equation}
It is sopposed that the function $x=x(T)$ defineds by the Eq.(38).

As we see from Eq.(42) $\Delta f$ is not only negative, as it have to be for
ordinary superconductors. Indeed for $T>T_{c1}$, where $T_{c1}=\sqrt{y_1}T_c$%
, $\Delta f$ is positive and according to Eq.(45) $c(t)$ and $d(T)$ will
change theirs signes. For ordinary superconductors it is impossible, but for
high-$T_c$ superconductors it takes place for temperatures $T>T_{c1}$. It
seems that for high-$T_c$ superconductors there are more than one critical
temperature. More work need to be done to understand thess new results. The
quanititative comparison of Eq. (\ref{42}) with the experiment will be given
in our next paper..

In closing I would like to thank S. G. Gevorgyan for discussions and
presentations of the experimatal work obtained in refs. \cite{2,3} which
stimulated this work. The author is also greatful to A. Mouradyan and M.
Hayrapetyan for the discussion of the results of this work.

\end{document}